\documentclass[11pt,a4paper]{article}
\usepackage[margin=25mm]{geometry}
\usepackage[T1]{fontenc}
\usepackage{lmodern,microtype,amsmath,amssymb,booktabs,array,cite,xcolor}
\usepackage[colorlinks=true,linkcolor=blue!50!black,citecolor=blue!50!black,urlcolor=blue!50!black]{hyperref}
\numberwithin{equation}{section}
\newcommand{\dd}{\mathrm d}
\newcommand{\tr}{\mathrm{tr}}
\newcommand{\str}{S_{\mathrm{tr}}}
\newcommand{\cO}{\mathcal O}
\newcommand{\cB}{\mathcal B}
\newcommand{\cE}{\mathcal E}
\newcommand{\trunc}{\mathrm{tr}}
\newcommand{\exact}{\mathrm{ex}}
\newcommand{\red}{\mathrm{red}}
\newcommand{\Ward}{\mathcal W}
\newcommand{\Amp}{\mathcal A}
\newcommand{\T}{\mathsf T}
\newcommand{\BF}{\mathsf B}
\newcommand{\FF}{\mathsf f}

\newif\ifreviewchanges
\reviewchangesfalse
\newcommand{\CMG}[1]{{\ifreviewchanges\color{red}\fi #1}}
\newenvironment{CMGblock}{\begingroup\ifreviewchanges\color{red}\fi}{\endgroup}
\title{Current Reconstruction and Higher Interactions\\
in Gauge Theory}
\author{Carolina Matt\'e Gregory\thanks{carolina.gregory@unb.br}\\[2mm]
\small Universidade de Bras\'ilia, Instituto de F\'isica,\\
\small 70910-900, Bras\'ilia, DF, Brasil\\
\small International Center of Physics,\\
\small Bras\'ilia, DF, Brasil}
\date{}
\begin{document}
\maketitle
\begin{abstract}
We revisit Deser's current reconstruction in first-order Yang--Mills
theory with an independent two-form. We fix the quadratic auxiliary
sector and the first gauge correction, and ask whether a prescribed
curvature interaction can be added without higher-order action terms.
For a compact semisimple gauge algebra, requiring the mixed Noether
identity to close without a $gh$ action term fixes the algebraic auxiliary
representative to the tangent representative, up to linear curvature and
dual-curvature terms on each simple ideal. For the analytic radial class
$U=\Phi(f\cdot f)$, with first nonlinear term $c_p(f\cdot f)^p$, no
allowed representative avoids a third-order action correction when all
second-order action coefficients vanish. A free-field scaling argument
inside a root $\mathfrak{su}(2)$ subalgebra establishes this result for
every compact semisimple gauge algebra. Ordinary covariantization and
an auxiliary-field shift give an explicit completion. \CMG{A different
free-field family establishes the same obstruction for the non-radial
quartic interaction $U=(f\cdot*f)^2$.} \CMG{For the radial member
$p=2$, the completion's}
seven-gluon contact term cancels a nonzero tree-level Ward contraction
at real nonsingular kinematics. \CMG{These results show how the
prescribed auxiliary data determine which interactions are required by
current reconstruction.}
\end{abstract}

\section{Introduction}

Deser's first-order construction of Yang--Mills theory is unusually
short: once an independent two-form is introduced, coupling the free
gauge field to its own current stops at cubic order \cite{Deser1970}.
Eliminating the auxiliary field restores the usual quartic vertex.  This
simple observation already suggests that finite self-coupling depends
on the variables and, more importantly here, on what one keeps fixed while
the current is reconstructed.

Seiberg--Witten (SW) maps provide another example.  A compact
Moyal action and its SW image are formally equivalent classical theories,
but their vertex expansions look very different
\cite{BarrocasPinzul,SeibergWitten,BarnichGrigorievHenneauxSW}.  Moreover,
the direct image of a current, the source appearing in the transformed
field equation, and the source obtained by varying the transformed action
need not agree off shell.  Their difference is controlled by the
variational chain rule and by improvements proportional to equations of
motion.  Related ambiguities in self-coupling and current improvement have
been discussed in Refs.~\cite{DeserHenneauxCurrents,Barcelo2021}; the Batalin-Vilkovisky (BV)
formulation of SW maps gives a systematic account of the corresponding
field redefinitions
\cite{BarnichGrigorievHenneauxSW,BarnichBrandtGrigorievSW}.

This paper asks a specific version of that question.  We keep the free
quadratic Hessian of the auxiliary field and the first localized gauge
transformation fixed, and vary the off-shell representative of a given
curvature interaction.  These are reconstruction conventions, not
consequences of gauge invariance.  If the auxiliary Hessian is also
allowed to change, a nonlinear gauge-invariant potential of the
independent two-form can absorb the interaction and the obstruction found
below disappears.  The result should therefore be read as a statement
about Deser-type reconstruction with fixed first-order data, not as a
no-go theorem for gauge-invariant actions in arbitrary variables.
\CMG{The fixed Hessian keeps the auxiliary equation linear in the
independent two-form. Section~\ref{auxiliary-comparison} compares this
choice explicitly with the nonlinear auxiliary potential and follows the
higher interactions generated when the free Hessian is restored.}

For a prescribed adjoint-invariant scalar $U(f)$ we consider
\[
 W_V=U(f)+(B-f)\cdot V(f),
\]
with $V$ an arbitrary smooth algebraic equivariant two-form.  This is the
full family that is affine in $B$ and reduces literally to $U$ on the free
auxiliary surface $B=f$.  The mixed Noether identity is quite restrictive when the $gh$ action
coefficient is set to zero.  In four dimensions, for a compact
semisimple algebra,
\[
 V=H+\sum_i(\alpha_i f_i+\beta_i *f_i),\qquad H=\frac{\partial U}{\partial f},
\]
with constant $\alpha_i,\beta_i$ on the simple ideals.  If the added
interaction starts at cubic field degree, the linear terms are absent and
the tangent representative $V=H$ is the only possibility.

The remaining question is whether another allowed representative, or a
higher correction to the gauge law, can remove the next action vertex.
For instance, for the radial class
\[
 U=\Phi(f\cdot f),\qquad
 \Phi(s)=c_p s^p+O(s^{p+1}),\qquad c_p\neq0,\quad p\geq2,
\]
the answer is no.  A one-parameter family of free fields separates the
possible representative correction from the Hessian contribution by
their powers of the field scale.  The calculation can be restricted to a
root $\mathfrak{su}(2)$ subalgebra and therefore applies to every compact
semisimple gauge algebra \cite{Hall2015}.  Arbitrary regular local higher
gauge corrections do not change the free-surface obstruction.

\CMG{We also test a non-radial quartic interaction, $U=(f\cdot*f)^2$.
The radial free-field family does not detect its third-order obstruction,
but adding a second polarization gives a nonzero residual for every
allowed representative. This extends the calculation beyond functions
of $f\cdot f$ while keeping the question restricted to an explicit
additional invariant.}

There is a familiar effective-field-theory interpretation of this family.
In four dimensions the $p=2$ member is an $F^4$ operator, at the same
algebraic order as the quartic terms in open-string and Born--Infeld
effective actions \cite{TseytlinNABI1997,TseytlinBIReview1999}.  We do not
identify $(F\cdot F)^2$ with the full non-Abelian Born--Infeld interaction, as
the latter contains a particular combination of Lorentz and color
contractions.  More generally, nonlinear functions of the curvature also
appear in the Yang--Mills deformations studied by Krasnov and by Cofano,
Fu and Krasnov \cite{Krasnov2015,CofanoFuKrasnov2015}.  Their construction
is chiral, while the family used here involves the full Lorentzian
two-form.  The comparison is meant only to place the interaction in a
standard higher-curvature setting.

\CMG{Section~\ref{data} fixes conventions and spells out the current
prescription, including the transformation of sources under field
redefinitions. Sections~\ref{completion}--\ref{third} construct the
completion, classify the allowed auxiliary representatives, and establish
the third-order obstruction for compact semisimple Yang--Mills theory.
Section~\ref{ward} gives the seven-gluon Ward check.
Appendix~\ref{affine} contains a separate star-gauge example and its
operator enumeration, while appendix~\ref{scattering-conventions}
records the scattering conventions and an independent local evaluation.
Throughout we work with classical formal deformations and tree amplitudes.}

\section{First-order fields and the current prescription}\label{data}

We work locally in four-dimensional Minkowski space with
$\eta=\operatorname{diag}(-1,1,1,1)$ and a compact semisimple real Lie
algebra $\mathfrak g=\bigoplus_i\mathfrak g_i$. The color pairing is
invariant and nondegenerate. Set
\begin{align}
 f_{ab}&=\partial_aA_b-\partial_bA_a,& K_{ab}&=[A_a,A_b],& C&=B-f,
 \label{fields}\\
 F&=f+gK,& L_0&=\tfrac14B\cdot B-\tfrac12B\cdot f,&
 L_{\mathrm p}&=-\tfrac12B\cdot K.\label{lagrangians}
\end{align}
The dot contracts color and both antisymmetric spacetime indices, with
no factor of $1/2$. The independent fields are $A,B$, and
\begin{equation}
 \delta S_0=\int\left(E_{0,A}^{b}\cdot\delta A_b+
                       \tfrac12 C\cdot\delta B\right),
 \qquad E_{0,A}^{b}=\partial_aB^{ab}.\label{euler}
\end{equation}
Actions are identified modulo divergences, and gauge parameters have
compact support. The free law and the prescribed first correction are
\begin{align}
 \delta_0A_a&=\partial_a\omega,&\delta_0B&=0,\notag\\
 \delta_1A_a&=[A_a,\omega],&\delta_1B&=[B,\omega].
 \label{gaugelaw}
\end{align}
Thus $\int(L_0+gL_{\mathrm p})$ is exactly invariant under
$\delta_0+g\delta_1$. Define
\begin{equation}
 X_\omega=\delta_0K,\qquad
 (X_\omega)_{ab}=[\partial_a\omega,A_b]+[A_a,\partial_b\omega].
 \label{X}
\end{equation}
In particular, $\delta_1f=[f,\omega]-X_\omega$ and
$\delta_1K=[K,\omega]$.

For a smooth algebraic adjoint-invariant scalar $U(f)$, define
\begin{equation}
 \delta U=H(f)\cdot\delta f,\qquad L_f[Y]=H'_f[Y].
 \label{gradient}
\end{equation}
The Hessian $L_f$ is self-adjoint. We consider the entire algebraic
family affine in $B$ with literal restriction $W|_{B=f}=U$,
\begin{equation}
 W_V=U(f)+(B-f)\cdot V(f),\qquad
 \str=\int(L_0+gL_{\mathrm p}+hW_V).\label{generalrep}
\end{equation}
Here $V$ is any smooth adjoint-equivariant two-form, with no gradient
assumption. Both $U$ and $V$ have no explicit coordinate dependence
or derivatives of $f$, and formal power series are also allowed. The
couplings $g,h$ are independent formal parameters and can have different
engineering dimensions. Unless a field or star-product expansion is specified, order means total
coupling degree in $g$ and $h$.

We prescribe the first gauge coefficients $g\delta_1$ and zero at order
$h$, and denote a mixed $gh$ gauge correction by $ghR_{11}$.
Interesting points to ask are whether the terms at $g^2$, $gh$ and $h^2$ can be assigned
entirely to the gauge law, and whether third-order action terms
must then appear. Higher gauge corrections
may act on both fields and contain arbitrarily many derivatives,
provided their differential order is finite at each coupling order.
They must be local and regular near the free stationary surface. We leave nonlocal operators or singular inverses of equations of motion
outside this problem.

It is useful to keep $U$, $V$ and the higher gauge corrections separate.
The first fixes the interaction on $B=f$, the second fixes its off-shell
continuation, and the last completes the symmetry.  A defect may be moved
into the gauge law at one order even though a new action vertex is forced
at the next.

The duality convention is $\epsilon_{0123}=+1$ and
$(*f)_{ab}=\epsilon_{abcd}f^{cd}/2$, so that $*^2=-1$ on two-forms.
With $\cE_i=f_{0i}$ and $\cB_i=\epsilon_{ijk}f_{jk}/2$, duality
acts as $(\cE,\cB)\mapsto(\cB,-\cE)$.

\subsection{What the current prescription fixes}

Integrating the $A$ equation determines the action only up to a
functional of the independent auxiliary field. This integration ambiguity
is distinct from the choice of $V$ in \eqref{generalrep}, namely, changing $V$
generally changes the off-shell $A$ current as well.

The restriction defines an inverse variational problem with specified
auxiliary data. For a local functional $S_W$, put $J_W=-E_A(S_W)$.
Scaling only $A$ gives the radial identity, modulo a boundary term,
\begin{equation}
 S_W[A,B]-S_W[0,B]
   =-\int_0^1\dd t\int A_b J_W^b[tA,B].\label{homotopy}
\end{equation}
Thus the $A$ current leaves a $B$-only functional undetermined. For a
term homogeneous of degree $n_A>0$ in $A$ at fixed $B$, the primitive
is $-\int A\cdot J_W/n_A$. Different homogeneous pieces require their
own factors. The associated $B$ Euler source is retained so that the
seed remains jointly variational and free gauge invariance of $W_V$
makes $J_W$ identically conserved.

\CMG{Keeping the $B$ Hessian equal to its free value restricts new
interactions to be affine in $B$ and keeps the auxiliary equation linear
in $B$, with the same algebraic coefficient as in the free theory.
This specifies the auxiliary response used in Deser's iteration.
Gauge invariance also permits a nonlinear response. For example, the
action}
\begin{equation}
 S_B=\int(L_0+gL_{\mathrm p}+hU(B))\label{Bonly}
\end{equation}
is exactly invariant under \eqref{gaugelaw} and linear in both
couplings. Its added $A$ Euler current vanishes, and a nonlinear $U(B)$
changes the auxiliary Hessian. For the tangent choice $V=H$,
\CMG{$U(B)-W_H=O(C^2)$ is proportional to products of the free
auxiliary equation and can be removed at first order by a local field
redefinition. That redefinition also induces higher-order terms.
Section~\ref{auxiliary-comparison} computes them explicitly and shows
how the same mixed vertices appear when the free auxiliary Hessian is
restored. Thus, the fixed-Hessian problem tests whether those terms can
be absorbed entirely into the higher gauge law while retaining a linear
auxiliary response.}

For fixed-current reconstruction, we also hold fixed the localized
adjoint law, the variational seed and its radial primitive, and compare
the recomputed current off shell modulo identically conserved
improvements, such as the current $\partial_aM^{ab}$ with $M^{ab}=-M^{ba}$. The strict absence of feedback is stronger than allowing an
extra gauge correction to cancel the localized variation.

\subsection{Field redefinitions and the reconstructed source}
\label{sources}

The distinction between an action and its current prescription is also
visible directly under a change of variables. Let
$\widehat\Phi=\mathcal M[\Phi]$, with $\Phi=(A,B)$, be a local,
near-identity, perturbatively invertible map with finitely many derivatives at each
formal order. Write $S_{\mathcal M}=\widehat S\circ\mathcal M$ and
let $\ell_{\mathcal M}$ be the Fr\'echet derivative of said map. Varying
the mapped action and integrating by parts gives
\begin{equation}
 \mathcal E_{\mathrm{act}}
   =\ell_{\mathcal M}^{\dagger}\mathcal E_{\mathrm{dir}},\qquad
 \mathcal E_{\mathrm{dir}}
   =\widehat{\mathcal E}\circ\mathcal M.
 \label{sourcechain}
\end{equation}
The collective Euler derivatives include the component pairing in
\eqref{euler}, and the dagger is the formal adjoint. Since the
near-identity operator $\ell_{\mathcal M}^{\dagger}$ is formally
invertible, the complete equation sets have the same formal solution
space. This is the usual variational relation underlying the BV
description of SW maps
\cite{BarnichGrigorievHenneauxSW,BarnichBrandtGrigorievSW}.

Now write $\widehat{\mathcal E}=\widehat{\mathcal K}-\widehat j$,
and choose a reference free operator $\mathcal K_0$ in the new
variables. Defining the sources by
$\mathcal E_{\mathrm{dir}}=\mathcal K_0-j_{\mathrm{eq}}$ and
$\mathcal E_{\mathrm{act}}=\mathcal K_0-j_{\mathrm{act}}$, we obtain
\begin{align}
 j_{\mathrm{eq}}
   &=\widehat j\circ\mathcal M
       -(\widehat{\mathcal K}\circ\mathcal M-\mathcal K_0),\notag\\
 j_{\mathrm{act}}-j_{\mathrm{eq}}
   &=-(\ell_{\mathcal M}^{\dagger}-1)\mathcal E_{\mathrm{dir}}.
 \label{threesources}
\end{align}
Thus, the image of the original current, the source in the transformed
equation and the source from the transformed action generally differ
off shell. Equation \eqref{sourcechain} compares the complete
dynamics, not identifying these three sources term by term.

For example, in Hermitian Moyal variables, the first-order connection
equation is
\begin{equation}
 \partial_\mu\widehat B^{\mu\nu}
   =\widehat j_{\mathrm D}^{\,\nu}
   =ig[\widehat A_\mu,\widehat B^{\mu\nu}]_\star.
 \label{moyalsource}
\end{equation}
A fixed SW map therefore gives
\begin{equation}
 j_{\mathrm{eq}}^{\,\nu}
   =\mathrm{SW}[\widehat j_{\mathrm D}^{\,\nu}]
      -\partial_\mu(\widehat B^{\mu\nu}[A,B]-B^{\mu\nu}).
 \label{swsource}
\end{equation}
The second term is an identically conserved improvement, whereas
the source derived from the action also contains the equation terms in
\eqref{threesources}. The independent $B$ field must be mapped as
such, replacing it by the curvature before the variation changes
this off-shell comparison.

If the fields, generator, pairing and interactions are transported
together, the complete theory is unchanged. Reusing the old current
prescription after changing only the action is a different operation.

\CMG{A finite star-gauge example in appendix~\ref{affine} makes the
distinction between strict current feedback and action truncation explicit.
Within its cubic operator sector, no nonzero representative has strictly
vanishing feedback, whereas a two-dimensional tangent family admits a
vanishing second-order action coefficient once the higher gauge law is
allowed to adjust. The common question is which interactions are required
when the auxiliary response and first gauge correction are held fixed.
The example is a separate reconstruction problem, with its own operator
sector and deformation parameter. The compact semisimple analysis below
starts directly from \eqref{generalrep} and uses neither an SW map from
that model nor its finite-dimensional classification.}

\CMG{We now keep the data in \eqref{generalrep} fixed and allow only
the higher gauge law to adjust.}

\section{Current feedback and higher interactions}\label{completion}

For the tangent representative the mixed $gh$ defect can be absorbed in
the gauge transformation, so no action term is needed at that order.  The
same truncated action, however, fails one order higher.  It is useful to
see both statements before turning to the general representative.

\subsection{Absorbing the mixed variation into the gauge law}

Consider first the tangent representative
\begin{equation}
 W_U=U+(B-f)\cdot H,
 \qquad \delta_fW_U=C\cdot L_f[\delta f].\label{tangent}
\end{equation}
Its mixed defect is $\delta_1W_U=-C\cdot L_f[X_\omega]$.
Because $C/2$ is the auxiliary Euler derivative, this variation can
be canceled without adding a mixed action term. The required law is
\begin{equation}
 \delta A_a=\partial_a\omega+g[A_a,\omega],\qquad
 \delta B=g[B,\omega]+2ghL_f[X_\omega].\label{exactlaw}
\end{equation}
With this law, no degree-two action terms are needed. The transformations have an
exactly closed off-shell algebra, as the invertible local variable
$\mathcal B=B+2hH(f)$ transforms as $\delta\mathcal B=g[\mathcal B,\omega]$.
The gauge parameter in the commutator is $g[\omega,\nu]$.

At this order, the defect is entirely proportional to the free auxiliary
equation and may be moved into $\delta B$.  No new action vertex is needed,
but the current prescription has changed.

The corrected gauge law makes the action invariant through total degree
two. To compare this with strict current feedback, we localize
the prescribed adjoint transformation. The resulting feedback can
be an identically conserved improvement precisely when
$\delta_1\int W_U$ vanishes for every local parameter. Taking the
independent $B$ Euler derivative, it gives $L_f[X_\omega]=0$ for every
$A,\omega$, which also makes the variation vanish. At fixed $f$, choose only
$\partial_a\omega=x$ and $A_b=y$ nonzero, with $a\ne b$. Then
$X_{ab}=[x,y]$ and its antisymmetric partner are the only components.
Since commutators span a semisimple algebra, these tensors span the
curvature space. Hence $L_f=0$ and $U$ is affine. Its invariant linear
coefficient vanishes because the algebra has no invariant covector,
so $U$ is constant on each connected domain. Every nonconstant $U$
in this tangent family therefore changes the prescribed current,
although the action needs no second-order coefficient once the
gauge law is corrected.

The same independent-$B$ variation for a general $V$ requires
$V'_f[X]=0$, and hence $V'=0$, if strict feedback is to vanish.
We will return to this condition after determining the allowed
representative freedom in Section~\ref{classification}.
The semisimple assumption is essential here, i.e., nonlinear interactions
confined to a central Abelian factor need not be detected.

\subsection{Covariantization and the first missing vertices}

To see where the higher action terms enter, apply ordinary
covariantization and an
auxiliary translation. Start with
\begin{equation}
 L_{\mathrm{cov}}=\tfrac14 B_c^2-\tfrac12 B_c\cdot F
      +h\{U(F)+(B_c-F)\cdot H(F)\},\qquad
 \delta B_c=g[B_c,\omega].\label{covaction}
\end{equation}
The change of variables $B_c=B+2h[H(f)-H(F)]$ starts at order $gh$
and preserves the prescribed first gauge correction. Completing the
square gives the exact identity
\begin{align}
 L_{\mathrm{cov}}
 &=\tfrac14(B_c-F+2hH(F))^2-\tfrac14F^2+hU(F)-h^2H(F)^2\notag\\
 &=L_0+gL_{\mathrm p}+hW_U+\mathcal R_U,\label{normalform}\\
 \mathcal R_U
 &=h\{U(F)-U(f)-H(f)\cdot(F-f)\}\notag\\
 &\quad+h^2\{H(f)^2-H(F)^2\}.\label{remainder}
\end{align}
The first braces in \eqref{remainder} start at $(gK)^2$, because the
linear term is already contained in the tangent representative; they give
the $g^2h$ vertex.  The second braces start at $gK$ and give the $gh^2$
vertex.  Both terms are therefore determined by the Taylor expansion of this
chosen covariant action.

The induced law is exactly \eqref{exactlaw}. In particular,
\begin{align}
 \mathcal R_U&=g^2hQ_{21}+gh^2Q_{12}+O_{\rm tot}(4),\notag\\
 Q_{21}&=\tfrac12 K\cdot L_f[K],&
 Q_{12}&=-2H\cdot L_f[K].\label{repair}
\end{align}
The construction is a choice of completion and invariant higher
interactions can be added. We use its explicit remainder rather than
claim uniqueness from the square identity.

Direct variation of the truncated action under \eqref{exactlaw}
gives the exact residual
\begin{equation}
 \delta\str=\int\{-g^2hK\cdot L_f[X_\omega]
                      +2gh^2H\cdot L_f[X_\omega]\}.\label{residual}
\end{equation}
The free variation of \eqref{repair} cancels the terms of degree three and the full remainder cancels the variation to all orders. For polynomial
$U$, this remainder is itself polynomial in the couplings, so a finite
completion exists. The obstruction concerns omission of the third-order
terms when the second-order action vanishes. Exact closure of the gauge
algebra alone does not make the truncated action invariant.

Eliminating the algebraic field gives
\begin{align}
 L_{\trunc,\red}&=-\tfrac14F^2+hU(f)+ghH(f)\cdot K-h^2H(f)^2,\notag\\
 L_{\exact,\red}&=-\tfrac14F^2+hU(F)-h^2H(F)^2.\label{reductions}
\end{align}
These expressions locate the required terms in an ordinary nonlinear
curvature theory. Auxiliary elimination and the square manipulation
use established machinery \cite{BBHGeneral1995,Krasnov2015}; the
representative restriction determines where the terms appear.

\begin{CMGblock}
\subsection{Comparison with a nonlinear auxiliary potential}
\label{auxiliary-comparison}

The action \eqref{Bonly} gives a useful comparison within the same
field content. Denote its independent two-form by $\widetilde B$.
At fixed $F$, its auxiliary equation has the formal branch
\begin{equation}
 b-F+2hH(b)=0,\qquad
 b(F,h)=F-2hH(F)+4h^2L_F[H(F)]+O(h^3).
 \label{nonlinear-auxiliary-branch}
\end{equation}
Substitution into its density gives
\begin{equation}
 L_{B,\red}(F,h)=-\tfrac14F^2+hU(F)-h^2H(F)^2
       +2h^3H(F)\cdot L_F[H(F)]+O(h^4).
 \label{nonlinear-auxiliary-reduced}
\end{equation}
Here the expansion is in $h$ at fixed $F$. This agrees with
$L_{\exact,\red}$ in \eqref{reductions} through order $h^2$, with the
first possible difference given by the displayed $h^3$ term.
Matching their first-order interaction does not identify their full
reduced actions.

The comparison can also be made off shell by restoring the free
auxiliary Hessian. Put $Y=\widetilde B-b$ and define
\begin{equation}
 \mathsf M(Y)=\mathbf1+4h\int_0^1\dd t\,(1-t)L_{b+tY},
 \qquad Z=\mathsf M(Y)^{1/2}Y.
 \label{auxiliary-normalization}
\end{equation}
Taylor expansion about the stationary point $b$ gives the exact identity
$L_B=\tfrac14Y\cdot\mathsf M(Y)Y+L_{B,\red}
=\tfrac14Z^2+L_{B,\red}$. The operator $\mathsf M$ is self-adjoint
under the two-form pairing and equals $\mathbf1+O(h)$. Its square
root is the formal branch about the identity, so this is a local,
perturbatively invertible algebraic change of auxiliary variable,
without a positivity assumption on the Lorentzian pairing.

Now set $Z=B-F+2hH(f)$. Adjoint equivariance makes $Z$ transform
covariantly, and the induced law for $B$ is precisely \eqref{exactlaw}.
Thus, the original first gauge correction is preserved. In these
variables, the same action becomes
\begin{align}
 L_B&=L_0+gL_{\mathrm p}+hW_U+\mathcal R_U+\Delta_U(F,h),
 \label{nonlinear-auxiliary-normalized}\\
 \Delta_U(F,h)&=L_{B,\red}+\tfrac14F^2-hU(F)+h^2H(F)^2\notag\\
 &=2h^3H(F)\cdot L_F[H(F)]+O(h^4).
 \label{auxiliary-completion-difference}
\end{align}
Its $B$ Hessian is now exactly the free one. In total coupling degree,
\begin{equation}
 \begin{split}
 L_B={}&L_0+gL_{\mathrm p}+hW_U+g^2hQ_{21}+gh^2Q_{12}\\
       &+2h^3H(f)\cdot L_f[H(f)]+O_{\rm tot}(4).
 \end{split}
 \label{auxiliary-comparison-third-order}
\end{equation}
There are no second-order action coefficients, while the mixed
third-order terms are exactly those of \eqref{repair}.
The additional $h^3$ coefficient is separately free-gauge invariant
and records the particular higher-order completion selected by
$S_B$.

For $U=(f\cdot f)^2$, the reduced expression is particularly simple,
\begin{equation}
 L_{B,\red}=-\tfrac14F^2+h(F^2)^2-16h^2(F^2)^3
                   +384h^3(F^2)^4+O(h^4).
 \label{nonlinear-auxiliary-quartic}
\end{equation}
Writing $s=f\cdot f$ and $t=f\cdot K$, the $h^2$ term contains
$-96gh^2s^2t$, the seven-field contact in \eqref{wardcontact}.
Thus, the action linear in $g,h$ already generates that contact
through its nonlinear auxiliary equation. Restoring the fixed
Hessian makes the corresponding $gh^2Q_{12}$ term explicit in
the independent-field action.

This normalization uses the established freedom in algebraic auxiliary
fields and the higher-order effects of perturbative field redefinitions
\cite{BBHGeneral1995,CriadoPerezVictoria2019}. It shows concretely how
nonlinear auxiliary response is exchanged for explicit higher
interactions. The further issue is necessity within the fixed-Hessian
class. The following sections exhaust its algebraic representatives
and show that, for the stated radial family and the non-radial quartic
interaction $U=(f\cdot*f)^2$, no choice of representative or regular local
higher gauge law avoids a third-order action term.
\end{CMGblock}

\section{Freedom in the auxiliary representative}\label{classification}

We now classify all $V(f)$ in \eqref{generalrep} for which the mixed
Noether identity can be satisfied without a $gh$ action term. On the free
surface, the problem reduces to a local constitutive condition for
Maxwell fields. The mixed identity can be solved with zero $gh$
action coefficient precisely for
\begin{equation}
 V(f)=H(f)+\sum_i\bigl(\alpha_i f_i+\beta_i {*f_i}\bigr),
 \label{allowed-representatives}
\end{equation}
where $f_i$ is the projection to the simple ideal $\mathfrak g_i$
and $\alpha_i,\beta_i$ are constants. The pure $g^2$ identity is
already satisfied by the first-order Yang--Mills action, and the
pure $h^2$ gauge coefficient can be zero. We therefore only need
to examine the mixed variation.

\subsection{The mixed Noether identity}

Write $T=V-H$ and let $V'_f[Y]$ denote the directional derivative
of $V$. Adjoint equivariance gives the exact identity
\begin{equation}
 \delta_1W_V=T\cdot X_\omega-C\cdot V'_f[X_\omega].
 \label{mixeddefect}
\end{equation}
The second term needs no symmetry assumption on $V'_f$.
Let $\Sigma_0$ be the prolonged free stationary surface
$C=0$, $\partial_af^{ab}=0$ and all their differential consequences.
Taking the Euler derivative with respect to the arbitrary parameter
$\omega$ before restriction to $\Sigma_0$, the mixed Noether identity
requires
\begin{equation}
 \left.E_\omega(T\cdot X_\omega)\right|_{\Sigma_0}=0.
 \label{necessary}
\end{equation}
Indeed, every regular $R_{11}S_0$ has vanishing parameter Euler
derivative on $\Sigma_0$.

By invariance of the color pairing,
\begin{equation}
 T\cdot X_\omega=2\langle[A_b,T^{ab}],\partial_a\omega\rangle,
 \qquad E_\omega(T\cdot X_\omega)=-2\partial_a[A_b,T^{ab}].
 \label{current}
\end{equation}
At a point, the value of $A_b$ is independent of the curvature and
its allowed first derivatives. Expanding the last expression gives
\begin{equation}
 \partial_a[A_b,T^{ab}]
 =\tfrac12[f_{ab},T^{ab}]+[A_b,\partial_aT^{ab}].
 \label{split}
\end{equation}
Varying each $A_b$ freely at fixed curvature jets in
\eqref{necessary} forces $\partial_aT^{ab}$ into the center
of $\mathfrak g$. Semisimplicity therefore implies
\begin{equation}
 \partial_aT^{ab}(f)=0
 \quad\text{on every local free Maxwell solution.}
 \label{constitutive}
\end{equation}
These are genuine local field data: a Maxwell two-form linear in
the coordinates and a local quadratic potential realize the
values and first derivatives used in this argument.

\subsection{The restriction from free Maxwell fields}

It remains to solve \eqref{constitutive}.  At a point, the spatial
derivatives of the electric and magnetic fields are arbitrary traceless
matrices, and the time derivatives are then fixed by Maxwell's equations.
These independent spatial jets are enough to determine the Jacobian of
$T(f)$.

Equation \eqref{constitutive} must hold for arbitrary free Maxwell
data, although $T$ depends only on the field strengths themselves.
For any finite collection $f^A$, a $C^2$ algebraic two-form without
explicit coordinate dependence can satisfy this condition only in
the form
\begin{equation}
 T^A=M^A{}_Bf^B+N^A{}_B{*f^B}+c^A,
 \label{algebraic-form}
\end{equation}
where $M,N$ and the two-forms $c^A$ are constant. The result follows
from the independence of the electric and magnetic spatial gradients,
without imposing Lorentz covariance on the function $T$.

For each input field, write $f=(\cE,\cB)$. Its equations are
\begin{equation}
 \nabla\cdot\cE=\nabla\cdot\cB=0,\qquad
 \partial_t\cE=-\nabla\times\cB,\qquad
 \partial_t\cB=\nabla\times\cE.
 \label{maxwell}
\end{equation}
At a point, the spatial derivative matrices of $\cE^B$ and
$\cB^B$ are arbitrary and independent traceless $3\times3$
matrices. The time derivatives are then fixed by
\eqref{maxwell}.

Let $(T_{\cE},T_{\cB})$ be the electric and magnetic components
of $T$. The time component of \eqref{constitutive} gives
$\nabla\cdot T_{\cE}=0$. The input derivative matrices are arbitrary
apart from their traces, so their contraction with the Jacobian of
$T_{\cE}$ can vanish only if that Jacobian is proportional to the
spatial identity:
\begin{equation}
 \frac{\partial(T_{\cE})^A_i}{\partial\cE^B_j}
     =M^A{}_B(f)\delta_{ij},\qquad
 \frac{\partial(T_{\cE})^A_i}{\partial\cB^B_j}
     =N^A{}_B(f)\delta_{ij}.
 \label{electricjac}
\end{equation}
Mixed partial derivatives force the coefficients to be constant.
For example, to differentiate $M^A{}_B$ with respect to any
$\cE^C_k$, choose $i\ne k$ and use
\begin{equation}
 \frac{\partial M^A{}_B}{\partial\cE^C_k}
 =\frac{\partial}{\partial\cE^B_i}
   \frac{\partial(T_{\cE})^A_i}{\partial\cE^C_k}=0.
\end{equation}
The other derivatives of $M,N$ vanish in the same way. Hence
$T_{\cE}=M\cE+N\cB+c_{\cE}$.

The spatial components of \eqref{constitutive} now give
\begin{equation}
 \nabla\times Z=0,\qquad
 Z=T_{\cB}-M\cB+N\cE.
\end{equation}
To see that $Z$ is constant, fix an input component $\cE^B_j$.
A spatial jet with only $\partial_i\cE^B_j\ne0$, $i\ne j$,
is traceless and allowed. The curl condition says that the
vector $\partial Z^A/\partial\cE^B_j$ is parallel to the $i$th
axis. There are two independent choices $i\ne j$, so this
vector is zero. The same argument applies to every magnetic
input. Thus $T_{\cB}=M\cB-N\cE+c_{\cB}$, giving
\eqref{algebraic-form}. Each term in that expression does satisfy
\eqref{constitutive}, by the Maxwell equations and Bianchi identities.

The appearance of $f$ and $*f$ agrees with the local two-form
cohomology of Maxwell theory \cite{BarnichBrandtHenneaux,AncoThe2005}.
That cohomology identifies expressions modulo exact terms. The
component calculation above fixes the literal algebraic expression
needed here, rather than dropping an exact term without examining it.

Since $T$ is adjoint-equivariant, its constant term takes values
in the center and therefore vanishes. Its linear coefficient
matrices commute with every adjoint action. For a compact
semisimple real algebra, such a linear map is a scalar multiple
of the identity on each simple ideal.
Equation \eqref{algebraic-form} consequently gives
\eqref{allowed-representatives} and no further algebraic dependence on $f$
is compatible with the mixed identity.

\subsection{The corrected gauge law and the remaining freedom}

Let $M$ and $N$ act as $\alpha_i$ and $\beta_i$ on the $i$th
simple ideal, and put $L_f=H'_f$. A mixed correction is
\begin{align}
 R_{11}A_a&=-2M[A_a,\omega],\notag\\
 R_{11}B&=2L_f[X_\omega]+4MX_\omega
           +2N{*X_\omega}-2M[f,\omega].
 \label{correction}
\end{align}
The corresponding curvature variation is
$R_{11}f=-2M([f,\omega]-X_\omega)$.
Substituting directly into the Lagrangian variations yields
\begin{equation}
 \delta_1W_V+R_{11}L_0
 =\sum_i\beta_i {*f_i}\cdot X_{\omega,i}.
 \label{boundary-remainder}
\end{equation}
Each term on the right is a divergence. In fact
\begin{equation}
 \partial_a[A_b,{*f^{ab}}]
 =\tfrac12[f_{ab},{*f^{ab}}]+[A_b,\partial_a{*f^{ab}}]=0
 \label{dual-current}
\end{equation}
identically, by the algebraic symmetry of the contraction and
the Bianchi identity. Hence all representatives in
\eqref{allowed-representatives}, including both linear exceptions, admit
the required mixed cancellation. The pure $g^2$ variation vanishes by
adjoint invariance of $L_{\mathrm p}$, and there is no pure
$h^2$ variation when its gauge coefficient is chosen zero.

For an added interaction that starts at cubic field degree, the
conditions are $U=O(f^3)$ and $V=O(f^2)$ near $f=0$, counting
$A$ and $B$ each with degree one. The linear terms in
\eqref{allowed-representatives} then disappear, leaving
\begin{equation}
 V=H,\qquad W_V=U+(B-f)\cdot H.
 \label{unique}
\end{equation}
The correction in \eqref{correction} reduces to the law already
used in \eqref{exactlaw}. Arbitrary higher field degrees and
mixtures are allowed by this argument. If quadratic interactions
are retained as well, the constants $\alpha_i,\beta_i$ describe
the full additional freedom.

We can also now return to strict current feedback. Its absence
requires $V'=0$, as found in section~\ref{completion}. A
nonconstant Hessian $H'_f$ cannot be canceled by the constant linear
terms in \eqref{allowed-representatives}. Quadratic and topological
exceptions require separate treatment, but they cannot remove the
feedback for the nonlinear family considered next.

\section{Third-order obstruction at fixed auxiliary Hessian}\label{third}

\subsection{The radial interaction}

We retain the four-dimensional Minkowski setting, compact semisimple
gauge algebra, and algebraic adjoint-equivariant representatives of
Section~\ref{data}. The auxiliary Hessian and first gauge correction
remain fixed, while arbitrary regular local higher gauge corrections are
allowed. We take $\Phi$ analytic near the vacuum, with
\begin{equation}
 U=\Phi(f\cdot f),\qquad
 \Phi(s)=c_p s^p+O(s^{p+1}),\qquad c_p\neq0,\quad p\geq2 .
 \label{radial-assumptions}
\end{equation}
Under these assumptions, setting all degree-two action coefficients to zero
is inconsistent with the third-order Noether identity, i.e., at least one
third-order action term is required.  We first show this for the
monomial $U=s^p$ and then extend the argument to the analytic class above.

The radial family has an explicit gradient and Hessian at every $p$,
which makes it possible to test the third-order identity uniformly in
the field degree.

We keep the two linear exceptions and test them on an explicit
$\mathfrak{su}(2)$ configuration.  The same configuration will later be
embedded in a general compact semisimple algebra.  Choose an orthonormal
basis
$[e_i,e_j]=\epsilon_{ijk}e_k$ and
\begin{equation}
 s=f\cdot f,\qquad U=s^p,\qquad p\in\mathbb N,\quad p\geq2.
 \label{family}
\end{equation}
The complete admissible family on this simple factor is
$V=H+T$, $T=\alpha f+\beta*f$. The parameters are constants
independent of the fields and of the independent formal
couplings.

With all degree-two action coefficients zero, the third-order
Noether residuals for \eqref{correction} are
$R_{21,\omega}=R_{11}L_{\mathrm p}$ and
$R_{12,\omega}=R_{11}W_V$, multiplying $g^2h$ and $gh^2$.
On $C=0$, write $P=\alpha I+\beta*$ to obtain
\begin{align}
 R_{12,\omega}|_{C=0}={}&2H\cdot L_f[X_\omega]
 +2T\cdot L_f[X_\omega]\notag\\
 &+4\alpha H\cdot X_\omega
 +2\beta H\cdot {*X_\omega}
 +2T\cdot P[X_\omega].
 \label{fullresidual}
\end{align}
This formula follows by varying before setting $C=0$:
$R_{11}W|_{C=0}=-T\cdot R_{11}f+(H+T)\cdot R_{11}B$.
Common adjoint contractions vanish by invariance.

\subsection{Freedom in the higher gauge transformations}

Every other solution of the mixed Noether identity differs
from \eqref{correction} by a free gauge symmetry. Completeness
of the Maxwell generators, retained upon adjoining an
algebraic auxiliary field, expresses the difference as a
free parameter redefinition plus a trivial symmetry,
\begin{equation}
 \Delta R\,\omega=R_0\Lambda[\omega]+M_\omega E_0.
 \label{trivial}
\end{equation}
Here $R_0$ denotes the free gauge generator, $E_0$ the collective free
Euler derivatives, and $M_\omega$ an equation-of-motion operator with the
appropriate antisymmetry after integrations by parts. This
is the standard completeness statement, see
Corollary~6.3 in Section~6.6 of \cite{BarnichBrandtHenneaux}, and auxiliary-field equivalence
is discussed in \cite{BBHGeneral1995}.

The first term in \eqref{trivial} annihilates $\int W_V$
by free gauge invariance. Its variation in $\int L_{\mathrm p}$
equals $-\delta_{1,\Lambda}S_0$ modulo a divergence. The second
term vanishes on the prolonged free surface, including after
the parameter Euler derivative is taken. The same reasoning
covers homogeneous gauge corrections at orders $g^2,h^2$.
Finally, every regular third-order gauge correction acts on
$S_0$ and has zero parameter Euler derivative on $\Sigma_0$.
Consequently, a nonzero restriction of $E_\omega R_{12,\omega}$
to $\Sigma_0$ is independent of all these gauge choices and
requires a higher action term.

\subsection{A family of free field configurations}

Let $\lambda$ be a freely variable constant field scale and put
\begin{equation}
 A_y=\lambda\bigl[(u+v)e_1+e_2\bigr],\quad
 A_t=A_x=A_z=0,\quad B=f,\qquad
 u=e^{-t+x},\quad v=e^{-t-x}.
 \label{scaled}
\end{equation}
This is an exact local free solution for every $\lambda$, with
\begin{equation}
 K=0,\qquad s=-8\lambda^2uv,\qquad f\cdot*f=0,
 \qquad (*f)\cdot X_\omega=0.
 \label{witness-data}
\end{equation}
No asymptotic boundary condition is intended: this is a
local test of the Noether identity. If a regular
neighborhood of the vacuum is required, $\lambda$ can be
arbitrarily small and nonzero.

For $U=s^p$, the first term of \eqref{fullresidual} is
\begin{equation}
 2H\cdot L_f[X_\omega]
 =8p^2(2p-1)s^{2p-2}f\cdot X_\omega.
\end{equation}
On \eqref{scaled}, the nonconstant remaining contribution
is $4\alpha p(2p+1)s^{p-1}f\cdot X_\omega$.
The $\beta$-dependent terms either vanish by
\eqref{witness-data} or multiply a current with zero
divergence on the free surface. Write
$\cO_{12}:=\left.E_\omega R_{12,\omega}\right|_{\Sigma_0}$.
Taking the parameter Euler derivative then gives only a third color component
\begin{align}
 \cO_{12}^{3}={}&-64p^2(p-1)(2p-1)8^{2p-2}
       \lambda^{4p-2}(uv)^{2p-2}(u+v)\notag\\
 &-16\alpha p(p-1)(2p+1)(-8)^{p-1}
       \lambda^{2p}(uv)^{p-1}(u+v).
 \label{anomaly}
\end{align}
At $t=x=0$, this becomes
\begin{align}
 \left.\cO_{12}^{3}\right|_0={}&
 -128p^2(p-1)(2p-1)8^{2p-2}\lambda^{4p-2}\notag\\
 &-32\alpha p(p-1)(2p+1)(-8)^{p-1}\lambda^{2p}.
 \label{origin}
\end{align}
The powers $\lambda^{4p-2}$ and $\lambda^{2p}$ are distinct for
$p\geq2$, and the coefficient of the former is nonzero. No constant
$\alpha$ can therefore make this expression vanish for all $\lambda$.
The result is independent of $\beta$.

For $p=2$ this reduces to
\begin{equation}
 \left.\cO_{12}^{3}\right|_0
 =-98304\lambda^6+2560\alpha\lambda^4.
 \label{F4}
\end{equation}
The polynomial cannot vanish on an open interval of field scales.
The scale is essential: a choice of $\alpha$ could cancel a single
unscaled evaluation, but cannot cancel the identity.

The $g^2h$ residual has zero parameter Euler derivative on
the same solution. In fact
\begin{equation}
 R_{11}L_{\mathrm p}
 =-\tfrac12(R_{11}B)\cdot K+\alpha B\cdot[K,\omega],
\end{equation}
and $K$ and all its derivatives vanish there.

Within the fixed-Hessian reconstruction problem, the action therefore
cannot stop at the first-order terms in this
$\mathfrak{su}(2)$ sector and $U=(f\cdot f)^p$, $p\geq2$, if all
second-order action coefficients are zero.
Equation \eqref{allowed-representatives} exhausts the admitted algebraic
representatives, while \eqref{origin} is nonzero as a polynomial in
the field scale for every choice of $\alpha,\beta$.  The freedom in the
higher gauge transformations does not alter this free-surface evaluation.

\subsection{Extension beyond \texorpdfstring{$\mathfrak{su}(2)$}{su(2)}}

The use of $\mathfrak{su}(2)$ above is a convenient realization of the
witness, not a restriction on the gauge algebra.  Let
$\mathfrak g=\bigoplus_i\mathfrak g_i$ be any compact semisimple real Lie
algebra and choose a simple ideal $\mathfrak g_i$.  A root of its
complexification determines a compact real
$\mathfrak h\simeq\mathfrak{su}(2)$ subalgebra
\cite{Hall2015}.  The invariant inner product of $\mathfrak g_i$
restricts to a nonzero multiple of the standard invariant form on
$\mathfrak h$.

Now take $A$, $B$ and the gauge parameter to be $\mathfrak h$-valued and
set all orthogonal color components to zero.  Since $\mathfrak h$ is
closed under the bracket, the free equations, the mixed gauge
correction and the Noether residual all remain in this subalgebra.  The
representative classification restricts to
\[
 V=H+\alpha_i f+\beta_i *f ,
\]
while $U=(f\cdot f)^p$ becomes the same power-law potential multiplied
by a nonzero normalization constant.  Such a normalization changes the
coefficients in \eqref{origin}, but not the two distinct powers
$\lambda^{4p-2}$ and $\lambda^{2p}$, nor the fact that the coefficient
of the former is nonzero.  Hence no constant $\alpha_i$ can cancel the
identity for all field scales, and $\beta_i$ remains irrelevant on the
same free witness.

The obstruction for the power-law interaction therefore holds for
\emph{every compact semisimple gauge algebra}, including all
$\mathfrak{su}(N)$ with $N\geq2$, not only for $\mathfrak{su}(2)$.  The $\mathfrak{su}(2)$ computation is
the minimal color sector in which the obstruction is detected.

\subsection{Beyond an exact monomial}

The monomial assumption can also be weakened without changing the
argument.  Let
\begin{equation}
 U=\Phi(s),\qquad
 \Phi(s)=c_p s^p+O(s^{p+1}),\qquad
 c_p\neq0,\quad p\geq2 ,
 \label{radial-general}
\end{equation}
with $\Phi$ analytic near the vacuum.  On the scaled free family,
the coefficient of $\lambda^{2p}$ in the parameter Euler derivative
is proportional to $\alpha_i c_p$ with the same nonzero factor as in
\eqref{origin}; an identity on an open interval of $\lambda$ therefore
first forces $\alpha_i=0$.  With this choice, the first surviving
coefficient is the $\lambda^{4p-2}$ term, proportional to $c_p^2$ and
to the nonzero first coefficient in \eqref{origin}.  Higher Taylor
coefficients in $\Phi$ enter only at higher powers of $\lambda$.
Consequently the same obstruction holds for the entire analytic radial
class \eqref{radial-general}.

The argument does not cover an arbitrary curvature invariant.  The
embedded $\mathfrak{su}(2)$ is not the difficulty; what matters is the
restriction of $U$ to that sector.  Extending the scaling argument requires control of the gradient and
Hessian contractions in \eqref{fullresidual}; agreement of the value of
$U$ on one free-field family is not sufficient. \CMG{The next subsection
demonstrates this point for $U=(f\cdot*f)^2$, using a different free-field
family. A classification of general Lorentz contractions and higher
colour invariants remains open.}

The same conclusion holds after $g=\zeta g_0$,
$h=\zeta h_0$, with $g_0h_0\ne0$, because the other
third-order residual vanishes on the same field configuration.
It is an obstruction to the specified action truncation, not a no-go
for gauge-invariant curvature theories or for auxiliary formulations in
which the quadratic $B$ Hessian is allowed to change. Nor does it make
the repairing coefficient unique modulo invariant additions.

\begin{CMGblock}
\subsection{A non-radial quartic interaction}\label{nonradial-quartic}

Consider the independent quartic invariant
\begin{equation}
 q=f\cdot*f,\qquad U=q^2,\qquad s=f\cdot f.
 \label{nonradial-invariant}
\end{equation}
With our Lorentzian conventions, the Hodge operator is self-adjoint
under the two-form pairing and satisfies $*^2=-1$. Consequently,
\begin{align}
 H&=4q*f,\qquad L_f[Y]=8(*f\cdot Y)*f+4q*Y,\notag\\
 2H\cdot L_f[X_\omega]
 &=-64sq(*f\cdot X_\omega)-32q^2(f\cdot X_\omega).
 \label{nonradial-Hessian}
\end{align}
On the original family \eqref{scaled}, both $q$ and
$*f\cdot X_\omega$ vanish identically. That family therefore gives
no obstruction for this interaction, even though its Hessian need
not vanish.

We retain the full representative freedom
$V=H+\alpha f+\beta*f$ in the $\mathfrak{su}(2)$ sector.
Substituting the gradient and Hessian above into
\eqref{fullresidual} gives
\begin{align}
 R_{12,\omega}|_{C=0}={}&
 \bigl[-32q^2-16\beta q+2(\alpha^2-\beta^2)\bigr]
       f\cdot X_\omega\notag\\
 &+\bigl[-64sq+40\alpha q-16\beta s+4\alpha\beta\bigr]
       *f\cdot X_\omega.
 \label{nonradial-full-residual}
\end{align}
Thus, the linear exceptions must be checked as well as the tangent
representative.

Using the same $u=e^{-t+x}$, $v=e^{-t-x}$ and colour basis as before,
replace \eqref{scaled} by
\begin{equation}
 \begin{gathered}
 A_y=\lambda\bigl[(u+v)e_1+e_2\bigr],\qquad
 A_z=\lambda(v-u)e_1,\\
 A_t=A_x=0,\qquad B=f.
 \end{gathered}
 \label{nonradial-witness}
\end{equation}
The new curvature equals $f_0+*f_0$, where $f_0$ is the curvature
of \eqref{scaled}. The free Maxwell equations and Bianchi identity
therefore hold for every $\lambda$. No duality symmetry of the
interacting theory is assumed. Direct contraction gives
\begin{equation}
 s=0,\qquad q=16\lambda^2uv,\qquad
 *f\cdot X_\omega=-f\cdot X_\omega.
 \label{nonradial-witness-data}
\end{equation}
In particular, $U$ varies along this family while $s$ remains zero,
so it cannot be a function of $s$ alone.

The constant multiple of $f\cdot X_\omega$ in the restricted
residual has zero parameter Euler derivative. For the remaining
terms, the current identity \eqref{current} gives
\begin{equation}
 \left.E_\omega(q^n f\cdot X_\omega)\right|_{\eqref{nonradial-witness}}
 =-4n\lambda^2q^n(u+v)e_3,\qquad n=1,2.
 \label{nonradial-current-evaluation}
\end{equation}
This follows from $\partial_t q=-2q$ and $\partial_xq=0$.
Combining \eqref{nonradial-full-residual}--\eqref{nonradial-current-evaluation}
and evaluating at $t=x=0$ yields
\begin{equation}
 \left.\cO_{12}^{3}\right|_0
 =131072\lambda^6+(5120\alpha+2048\beta)\lambda^4,
 \qquad \cO_{12}^{1}=\cO_{12}^{2}=0.
 \label{nonradial-obstruction}
\end{equation}
The coefficient of $\lambda^6$ is nonzero and cannot be cancelled
by any constant $\alpha,\beta$ on an open interval of field scales.
The freedom in higher gauge transformations drops out on the
prolonged free surface by the same argument used above. Hence a
third-order action coefficient is required for $U=q^2$ whenever
all second-order action coefficients vanish.

Here $K_{yz}=\lambda^2(u-v)e_3$ need not vanish, but the other
mixed residual still has zero parameter Euler derivative. Indeed,
$K\cdot L_f[X_\omega]=K\cdot*X_\omega=0$, and its only remaining
density is $R_{21,\omega}=-2\alpha K\cdot X_\omega$. This contains
only $y,z$ derivatives of the parameter, with coefficients depending
only on $t,x$, so $E_\omega R_{21,\omega}=0$ on this family.
The conclusion therefore also survives the specialization
$g=\zeta g_0$, $h=\zeta h_0$ with $g_0h_0\ne0$.

The root-subalgebra embedding applies unchanged. The gradient and
Hessian preserve the chosen $\mathfrak{su}(2)$ sector, and the
nonzero normalization of its invariant pairing cannot remove the
$\lambda^6$ coefficient. Thus, this additional obstruction holds
for every compact semisimple gauge algebra under the same
fixed-Hessian assumptions.

For the tangent representative, the explicit repair
\eqref{repair} specializes to
\begin{align}
 Q_{21}&=4(*f\cdot K)^2+2q(K\cdot*K),\notag\\
 Q_{12}&=64sq(*f\cdot K)+32q^2(f\cdot K).
 \label{nonradial-repair}
\end{align}
Their free variations cancel the corresponding third-order
residuals. This provides a further worked interaction and its
completion, while leaving combinations of independent quartic
invariants and higher colour contractions outside the present
obstruction result. The scattering calculation below remains the
separate check for $U=s^2$.
\end{CMGblock}

\section{Seven-gluon scattering and the missing contact}\label{ward}

The local free solution in Section \ref{third} tests a local Noether
identity. It is not a scattering configuration. We now give an
independent on-shell diagnostic of this truncation for the tangent
representative with
$\mathfrak{su}(2)$ and $U=s^2$, where $s=f\cdot f$ and $t=f\cdot K$.
The algebraic equation is $B=F-8hsf$, and \eqref{reductions} becomes
\begin{align}
 L_{\trunc,\red}&=-\tfrac14F^2+hs^2+4ghst-16h^2s^3,\label{wardtrunc}\\
 L_{\exact,\red}&=-\tfrac14F^2+h(F^2)^2-16h^2(F^2)^3.\label{wardexact}
\end{align}
The mixed terms in the reduced truncation are induced by eliminating
$B$ and they do not contradict the absence of degree-two coefficients
in the independent-field action. At order $gh^2$, the difference is
the seven-field, five-derivative contact
\begin{equation}
 Q_{12}=-96s^2t.\label{wardcontact}
\end{equation}
The $g^2h$ difference cannot contribute to the independent coefficient
$gh^2$. There are no external auxiliary states.

The scattering calculation is only an on-shell check of the local
obstruction.  We compare the exact and truncated reduced actions, which
isolates the missing seven-point contact before any diagrams are summed.

\subsection{From the action difference to the Ward defect}

The on-shell gluon Ward identity is the vanishing of an amplitude
when one polarization is replaced by its momentum, with the other
polarizations physical \cite{BernWard2014}, and no soft limit is needed.
Let $\Ward_{7,\trunc}^{[1,2]}$ denote this contraction on leg 7 for
the full color amplitude at coupling $gh^2$. We remove a common
Fourier/Feynman phase, use multilinear density coefficients with
$\partial\to k$ as vertices, and use $-\eta^{\mu\nu}\delta^{ab}/p^2$
for each internal edge. Appendix \ref{scattering-conventions} fixes these
conventions explicitly.

For six physical legs define
\begin{align}
 A^\mu(z)&=\sum_{i=1}^6z_i\varepsilon_i^\mu c_i,&
 f^{\mu\nu}(z)&=\sum_{i=1}^6z_i
   (k_i^\mu\varepsilon_i^\nu-k_i^\nu\varepsilon_i^\mu)c_i,\notag\\
 X_7^{\mu\nu}(z)&=[k_7^\mu c_7,A^\nu(z)]
                   +[A^\mu(z),k_7^\nu c_7].\label{wardwaves}
\end{align}
Plane-wave factors are understood, and $\sum_i k_i=0$.
The missing contact determines the entire Ward defect. For
\eqref{wardtrunc}--\eqref{wardexact}, we find
\begin{equation}
 \Ward_{7,\trunc}^{[1,2]}
 =96[z_1\cdots z_6]\,(f(z)^2)^2f(z)\cdot X_7(z)
 =-\Ward_7[Q_{12}].\label{wardidentity}
\end{equation}
To obtain this relation, compare the diagrams of the two actions.
At this order the actions have identical exchange diagrams and
differ only by \eqref{wardcontact}. An additional positive-coupling
vertex attached to that contact would exceed $gh^2$. Gauge invariance
of \eqref{wardexact} makes the full Ward contraction zero.
When $\varepsilon_7=k_7$, the leg's linear curvature vanishes and
it can appear in the contact only through $K$, producing $X_7$.
This gives \eqref{wardidentity}.

For $H=4sf$ and $L_f[Y]=8(f\cdot Y)f+4sY$, the local density on
the right is exactly $2H\cdot L_f[X_\omega]$ in
\eqref{residual}. Equation \eqref{wardidentity} identifies its
on-shell multilinear evaluation with the full exchange Ward defect.
More generally, for an invariant homogeneous $U$ of degree $m\geq3$,
the same action comparison gives
\begin{equation}
 \Ward_{2m-1,\trunc}^{[1,2]}
 =[z_1\cdots z_{2m-2}]\,2H(f(z))\cdot L_{f(z)}[X_{2m-1}(z)].
 \label{generalward}
\end{equation}
This does not assert nonvanishing for every $U$. The explicit
nonvanishing result below is for $U=s^2$.

\subsection{Real nonsingular kinematics and the full tree sum}

Use contravariant outgoing momenta in an arbitrary momentum unit
$\mu=1$, with legs 1 and 2 incoming physically. All external momenta
are nonzero. The color basis satisfies $[e_i,e_j]=\epsilon_{ijk}e_k$.
\begin{center}
\begin{tabular}{cccc}
\toprule
Leg & $k_i/\mu$ & $\varepsilon_i$ & Color \\
\midrule
1 & $(-10,-6,0,-8)$ & $(0,0,1,0)$ & $e_1$ \\
2 & $(-10,6,0,8)$ & $(0,0,1,0)$ & $e_1$ \\
3 & $(3,3,0,0)$ & $(0,0,1,0)$ & $e_1$ \\
4 & $(4,0,4,0)$ & $(0,1,0,0)$ & $e_1$ \\
5 & $(5,-3,-4,0)$ & $(0,4,-3,0)$ & $e_1$ \\
6 & $(4,0,0,4)$ & $(0,1,0,0)$ & $e_2$ \\
7 & $(4,0,0,-4)$ & $(0,1,0,0)$ & $e_3$ \\
\bottomrule
\end{tabular}
\end{center}
These data obey $k_i^2=0$, $k_i\cdot\varepsilon_i=0$ and momentum
conservation. All 56 two-leg and three-leg channel invariants are
nonzero. The relevant vertex densities, with couplings stripped, are
\begin{equation}
 Y_3=-\tfrac12t,\quad V_4=s^2,\quad V_5=4st,
 \quad W_6=-16s^3,\quad W_7=-96s^2t,\label{wardvertices}
\end{equation}
with couplings $g,h,gh,h^2,gh^2$, respectively. At fixed coupling
$gh^2$, there are four exchange topologies, and their labeled counts can be read off
without generating graphs. For $Y_3$ joined to $W_6$, choose the two external legs
attached to the cubic vertex, giving $\binom72=21$. For $V_4$ joined to $V_5$, choose
the three external legs on the quartic vertex; the two ends are distinguishable, so the
count is $\binom73=35$. For a cubic vertex between two identical $V_4$ vertices, choose
the single external leg at the cubic vertex and split the remaining six legs into two
unordered triples, giving $7\binom63/2=70$. Finally, when a $V_4$ vertex lies between a
$Y_3$ and another $V_4$, choose the two external legs on $Y_3$ and then three of the
remaining five for the terminal $V_4$, giving $\binom72\binom53=210$. These four
numbers sum to 336 exchange graphs. The Yang--Mills quartic vertex has coupling $g^2$
and therefore cannot enter this coefficient, while ghost lines cannot form an all-gluon
tree.

Replacing $\varepsilon_7$ by $k_7$, exact rational arithmetic gives
the following sums. Graph counts include diagrams whose numerators
vanish at this point.
\begin{center}
\begin{tabular}{lrr}
\toprule
Contribution & Graphs & Ward coefficient at $\mu=1$ \\
\midrule
$Y_3$ joined to $W_6$ & 21 & $2\,812\,280\,832$ \\
$V_4$ joined to $V_5$ & 35 & $879\,815\,950\,336/665$ \\
$Y_3$ between two $V_4$ & 70 & $-8\,042\,119\,168/133$ \\
$V_4$ between $Y_3$ and $V_4$ & 210 & $-839\,605\,354\,496/665$ \\
\midrule
All exchanges: truncated action & 336 & $2\,812\,280\,832$ \\
Contact $W_7$ & 1 & $-2\,812\,280\,832$ \\
Completed action & 337 & $0$ \\
\bottomrule
\end{tabular}
\end{center}
The last three classes contain $35+70+210=315$ graphs.  They are the
order-$gh^2$ exchanges generated by the gauge-invariant $F^4$ family and
their Ward sum vanishes on its own, as expected for
$-F^2/4+h(F^2)^2$.  The remaining 21 graphs join the Yang--Mills cubic
vertex to the $h^2$ six-point interaction $-16s^3$; their Ward sum is
canceled by the single contact $-96gh^2s^2t$.  This separation is a useful
check on the direct diagrammatic sum.

The physical meaning can also be stated as invariance under a change
of polarization representative. At $\mu=1$, both $\varepsilon_7$
and $\varepsilon_7+k_7$ are transverse and have the same norm, but
\begin{align}
 \Amp_{7,\trunc}^{[1,2]}(\varepsilon_7+k_7)
       -\Amp_{7,\trunc}^{[1,2]}(\varepsilon_7)&=2\,812\,280\,832,\notag\\
 \Amp_{7,\exact}^{[1,2]}(\varepsilon_7+k_7)
       -\Amp_{7,\exact}^{[1,2]}(\varepsilon_7)&=0.\label{shiftward}
\end{align}
The completed physical coefficient itself is
$-1\,398\,336\,978\,944/21\,945$. The Ward contraction scales as
$\mu^6$ and the physical coefficient as $\mu^5$. Thus the local
obstruction has a nonzero on-shell consequence at a nonsingular
physical point. Continuity gives nonzero values in a neighborhood
of this point within the allowed kinematic data.

\section{Discussion}\label{discussion}

The result concerns a specific reconstruction problem.  In a first-order
current construction, one specifies more than the final curvature interaction:
one also chooses the independent fields, the off-shell representative and
the way the rigid symmetry is localized.  These choices are not preserved
term by term under a general field redefinition, even when the complete
classical theory is. \CMG{The source relations in
section~\ref{sources} make this dependence explicit.}

\CMG{For compact semisimple Yang--Mills theory, the mixed Noether
identity allows us to}
classify the algebraic representatives.  Apart from linear $f$ and $*f$
terms on each simple ideal, $V$ is fixed to be the gradient
$H=\partial U/\partial f$.  No gradient ansatz for $V$ was used.  For
$U=\Phi(f\cdot f)$, with first nonlinear term $c_p(f\cdot f)^p$, the
free-field scaling argument then forces a nonzero third-order action
coefficient.  The explicit witness is $\mathfrak{su}(2)$-valued, but a root
$\mathfrak{su}(2)$ subalgebra embeds it in any compact semisimple gauge
algebra. \CMG{The same conclusion holds for the non-radial quartic
interaction $U=(f\cdot*f)^2$. The original free-field configuration
gives a vanishing residual, but the modified configuration in
section~\ref{nonradial-quartic}
gives a nonzero obstruction for every allowed representative.}

The seven-gluon calculation gives an on-shell test of the local
residual.  For $U=(f\cdot f)^2$, the truncated
reduced action has a nonzero longitudinal tree amplitude at a real,
nonsingular kinematic point.  The contact term supplied by the covariant
completion cancels it exactly.  This does not make the fixed-Hessian
convention physical; it only shows that, once that convention is adopted,
omitting the required vertex has an ordinary on-shell consequence.

The result covers the non-chiral analytic class
$\Phi(F\cdot F)$, whose first nonlinear Taylor coefficient is
$c_p(F\cdot F)^p$ with $p\geq2$, \CMG{together with the additional
quartic invariant $(F\cdot*F)^2$. This example shows why the
free-field configuration must be chosen to test the gradient and
Hessian of the interaction. It does not establish the obstruction
for arbitrary combinations of these invariants or for higher
colour contractions, which remain separate extension questions.}

\CMG{The nonlinear auxiliary formulation makes the role of the
restriction explicit. Although $S_B$ is linear in the couplings,
eliminating its auxiliary field generates the same order-$gh^2$
contact considered here. Restoring the free auxiliary Hessian instead
produces explicit mixed third-order vertices, together with the
invariant completion term displayed at order $h^3$.
This redistribution is consistent with the established treatment of
field redefinitions at higher orders \cite{CriadoPerezVictoria2019}.
The additional result is the exhaustive restriction on representatives
and the necessity of a higher action term for the stated radial family
and the additional quartic invariant.
It concerns reconstruction with a fixed linear auxiliary response and
the prescribed first gauge law, while consistent local gauge
deformations remain available when the higher interactions are retained.}

\appendix

\begin{CMGblock}
\section{An affine star-gauge example of current reconstruction}\label{affine}

This appendix gives a separate example of the distinction between
strict current feedback and action truncation discussed in
section~\ref{sources}. Restricting to a finite cubic operator sector
allows the representative quotient and both reconstruction conditions
to be computed explicitly. We specify the star product and the
independent auxiliary field first, then give the operator enumeration
and the local calculations supporting the result. These statements
concern this sector and do not establish an SW-map relation to the
radial interaction of section~\ref{third}.
\end{CMGblock}

\begin{CMGblock}
\subsection{Star product and reconstruction conditions}\label{affine-main}
\end{CMGblock}

On the regular patch $x>0$, set $u=\log x$ and use the commuting frame
\begin{equation}
 D_a=(\partial_t,x\partial_x,\partial_y,\partial_z),\qquad
 \dd^4X=\frac{\dd t\,\dd x\,\dd y\,\dd z}{x},\qquad \tau^{12}=1.
 \label{affineframe-main}
\end{equation}
The frame indices are $a=0,1,2,3$, corresponding to $(t,u,y,z)$, and the
metric is Minkowski in these coordinates. With
$P(v,w)=\tau^{ij}D_ivD_jw$ and
\begin{equation}
 v\star w=v\exp\!\left(\frac{i\kappa}{2}\overleftarrow D_i
 \tau^{ij}\overrightarrow D_j\right)w,
 \label{affinestar-main}
\end{equation}
the measure is cyclic.  For scalar fields at fixed $g$, the star
commutator has no $\kappa^2$ term and
$F=f+\epsilon K+O(\kappa^3)$, where $\epsilon=g\kappa$,
$f_{ab}=D_aA_b-D_bA_a$ and $K_{ab}=P(A_a,A_b)$.  The first-order
interaction and localized gauge correction are
\begin{equation}
 L_{\mathrm p}=-\frac12B\cdot K,\qquad
 \delta_1A_a=-P(\omega,A_a),\qquad
 \delta_1B_{ab}=-P(\omega,B_{ab}).
 \label{affinelaw-main}
\end{equation}
This is a regular commuting-frame star product, not a statement across the
rank-changing locus or for an arbitrary Poisson tensor.

We consider cubic densities that are affine in $B$ and have engineering
weight six, assigning weights $1$ to $A$ and $D$, $2$ to $B$, and $0$
to the fixed tensor $\tau$. The coefficients are constant, there is
exactly one $\tau$, contractions use only $\eta$ and $\tau$, and no
explicit coordinate dependence is allowed. Derivatives of $B$ are moved
onto the other fields by integrating by parts. Among these densities,
those invariant under the free gauge law modulo a divergence form a
six-dimensional space, also modulo divergences,
\begin{align}
 v&=\sum_{i=1}^4\lambda_i V_i+\rho_1U_1+\rho_2U_2,\notag\\
 V_1&=(\tau\cdot B)f^2,&
 V_2&=(\tau\cdot f)(B\cdot f),\notag\\
 V_3&=\tr(\T\BF\FF^2),&
 V_4&=\tr(\T\FF\BF\FF),\label{affinebasis-main}\\
 U_1&=(\tau\cdot f)f^2,&
 U_2&=\tr(\T\FF^3)=\tau^{ab}f_{ac}f_{bd}f^{cd}.
 \notag
\end{align}
Here $\T_a{}^b=\tau_{ac}\eta^{cb}$ and, more generally,
$\mathsf X_a{}^b=X_{ac}\eta^{cb}$ for any two-form $X$.  The curvature structures
$U_1,U_2$ are of the familiar type that appears at first order in
noncommutative gauge theory, while the $V_i$ retain the independent
auxiliary representatives. See, for example, \cite{Latas2007}.
\CMG{The enumeration in subsection~\ref{affine-enumeration} shows
that these six operators exhaust the stated sector.}

Four directions in \eqref{affinebasis-main} are free-equation shifts of
$B$ and only change the continuation away from $B=f$.  The field-redefinition quotient
is therefore represented by
\begin{equation}
 a=\lambda_1+\lambda_2+\rho_1,\qquad
 b=\lambda_3+\lambda_4+\rho_2,
 \label{affinequotient-main}
\end{equation}
and is exactly two-dimensional. The reconstruction conditions are imposed
on representatives of these classes, since strict current feedback is
not invariant under the field redefinitions used in forming the quotient.
If the first gauge law
\eqref{affinelaw-main} is held fixed, and one demands both vanishing
second-order action and strictly vanishing current feedback modulo an
identically conserved improvement, the only solution is $v=0$.  Thus, a
nonzero first-order interaction inevitably feeds back into the fixed
current prescription.

Allowing the regular second gauge correction required by the Noether
identity changes the answer.  The complete family for which the action may
still satisfy $S_2=0$ is
\begin{equation}
 v=aW_1+bW_2,\qquad
 W_1=V_1+2V_2-2U_1,\qquad
 W_2=2V_3+V_4-2U_2.
 \label{affinetangent-main}
\end{equation}
For $U=aU_1+bU_2$ and $H_U=\partial U/\partial f$, these combinations are
precisely
\begin{equation}
 v=U+(B-f)\cdot H_U=B\cdot H_U-2U.
 \label{affinetangentform-main}
\end{equation}
Here $v$ enters the action as $\epsilon v$, alongside
$\epsilon L_{\mathrm p}$, and $S_2$ denotes the coefficient of
$\epsilon^2$. With $X_\omega=\delta_0K$, the choice
\begin{equation}
 \delta_2A=0,\qquad \delta_2B=2H'_U[X_\omega]
 \label{affinegauge2-main}
\end{equation}
gives $\delta_1v+\delta_2L_0=-P(\omega,v)$, which integrates to zero by
cyclicity.  Thus, strict current feedback and second-order action truncation
select different subspaces even in this explicit star-gauge example.

Conversely, every class in \eqref{affinebasis-main} has the second-order
variational completion
\begin{equation}
 L_{2,v}=H_v\cdot K,\qquad H_v=\frac{\partial v}{\partial f}\bigg|_B.
 \label{affinecompletion-main}
\end{equation}
Let $N_v^b$ be the feedback current defined in
\CMG{\eqref{affineanomaly}}. If $G_v=(\partial_fH_v)[K]$ and
$J_{2,v}=-E_A(H_v\cdot K)$, direct variation gives
\begin{align}
 N_v^b-J_{2,v}^b&=D_aM_v^{ab},\notag\\
 M_v^{ab}&=2\bigl\{(\tau^{bj}H_v^{ac}-\tau^{aj}H_v^{bc})D_jA_c
                         -G_v^{ab}\bigr\},\qquad M_v^{ab}=-M_v^{ba}.
 \label{affinecurrent-main}
\end{align}
The feedback therefore has an explicit variational repair modulo an
identically conserved improvement.

\CMG{The affine calculation thus separates strict current feedback from
second-order action truncation in the stated operator sector. The
calculations below establish the completeness of this sector and the
two reconstruction conditions directly.}

The square identity \CMG{\eqref{normalform}} also holds for an invariant functional with an
equivariant gradient and formally self-adjoint Hessian under a cyclic
pairing. A fixed associative star product with commuting derivations
and a cyclic trace is an example. Alternative twist-based constructions
may instead deform the differential calculus together with the star
product, retaining gauge-covariant structures for a wider class of
noncommutative deformations \cite{BorsatoMeier2026,BorsatoMeierPerspective2026}.
This functional extension does not extend the finite-dimensional
semisimple argument to a general star algebra or to arbitrary Poisson
structures.

\subsection{Completeness of the operator sector}\label{affine-enumeration}

With the affine-in-$B$ restriction of \CMG{subsection~\ref{affine-main}},
move derivatives off $B$ by integration by parts and generate the five
tensor words
\begin{equation}
 \tau B(DA)(DA),\quad \tau BA(D^2A),\quad
 \tau(DA)^3,\quad \tau A(DA)(D^2A),\quad \tau A^2(D^3A).
 \label{affinewords}
\end{equation}
Each word has eight indices. Pairing them with four metrics gives
$(8-1)!!=8!/(2^4 4!)=105$ contractions before tensor symmetries are imposed. The first
two words contain one $B$ and therefore contribute $2\times105=210$ raw contractions, and the remaining three are pure-$A$ words and contribute $3\times105=315$, for a total of
525. The overcomplete list avoids assuming from the outset that every
representative can be written only in terms of curvatures.

Antisymmetry and commuting derivatives are then imposed exactly. The ``density rank''
column below is the dimension left after these algebraic identities. Applying the Euler
map removes total divergences in this positive-field-degree polynomial class and the
``Modulo $\dd$'' column is the resulting quotient dimension. Finally, the parameter
Euler derivative of the free gauge variation gives a linear map on that quotient. Its
rank is shown in the next column, so the last column is simply the dimension of its
kernel. All row reductions are over the rationals. The calculation gives
\begin{center}
\begin{tabular}{lrrrrr}
\toprule
Sector & Generated & Density rank & Modulo $\dd$ & Gauge rank & Kernel\\
\midrule
One $B$ & 210 & 24 & 24 & 20 & 4\\
Pure $A$ & 315 & 37 & 11 & 9 & 2\\
\midrule
Total & 525 & 61 & 35 & 29 & 6\\
\bottomrule
\end{tabular}
\end{center}
Thus the one-$B$ sector has a four-dimensional gauge-invariant quotient and the pure-$A$
sector has a two-dimensional one. The four $V_i$ and two $U_j$ in
\eqref{affinebasis-main} have independent Euler images in these respective kernels and
therefore exhaust the homogeneous freedom.
The raw generation also excludes missing $AD^2A$ representatives; counting
curvature contractions alone would not establish that assertion.

Writing $V_i=B\cdot Q_i(f)$ gives
$V_i-U_{\sigma(i)}=(B-f)\cdot Q_i$, with
$\sigma(1)=\sigma(2)=1$ and $\sigma(3)=\sigma(4)=2$.
The shifts $B\mapsto B+2\epsilon c_iQ_i$ remove four free-equation
directions and preserve the first gauge correction because
$\delta_0Q_i=0$.  The quotient coordinates are therefore those in
\eqref{affinequotient-main}.  A certificate of their independence uses
contravariant complex incoming momenta
\begin{equation}
 k_1=(1,1,0,0),\quad k_2=(0,0,1,i),\quad
 k_3=(-1,-1,-1,-i),
\end{equation}
with $e_1=(0,0,1,0)$ and $e_2=(1,0,0,0)$ for both choices, and
$e_3=(1,1,0,0)$ or $(1,0,1,0)$. They are null, transverse and conserve
momentum. The multilinear vertices of $(U_1,U_2)$ are the two rows of
\begin{equation}
 \begin{pmatrix}8&6\\-16&-4\end{pmatrix},\qquad \det=64.
\end{equation}
Free-equation terms and divergences vanish on these states. This analytic
three-point calculation establishes independence modulo local redefinitions and 
it is not a claim about real massless three-body decay.

\subsection{Strict feedback and the tangent kernel}\label{affine-kernels}

For $H_v=\partial v/\partial f$ at fixed $B$, the feedback current under
\eqref{affinelaw-main} has representative
\begin{equation}
 N_v^i=2\tau^{ij}D_a(H_v^{ab}D_jA_b),\qquad
 \mathcal A(v)=-D_iN_v^i.
 \label{affineanomaly}
\end{equation}
Six independent coefficients in this off-shell polynomial are
\begin{center}
\begin{tabular}{lc}
\toprule
Jet monomial & Coefficient in $\mathcal A(v)$\\
\midrule
$A_{0,013}A_{3,2}B_{12}$ & $-8\lambda_1$\\
$A_{0,1}A_{1,2}B_{03,23}$ & $-4\lambda_2$\\
$A_{0,01}A_{3,1}B_{23,2}$ & $-\lambda_3$\\
$A_{0,1}A_{1,3}B_{03,22}$ & $-2\lambda_4$\\
$A_{0,013}A_{2,1}A_{3,2}$ & $-8\rho_1$\\
$A_{0,01}A_{2,13}A_{3,2}$ & $-2\rho_2$\\
\bottomrule
\end{tabular}
\end{center}
These six monomials isolate the six unknown coefficients.  The resulting
$6\times6$ matrix is diagonal up to the displayed nonzero factors and has
determinant 1024.  Hence, $\mathcal A(v)=0$ identically only for $v=0$.
No nonzero representative in this sector has strictly vanishing feedback
modulo an identically conserved improvement.

Allowing a regular second gauge correction gives the two-dimensional kernel
\eqref{affinetangent-main}. To determine it, evaluate
$E_\omega(\delta_1v)$ on prolonged free solutions. With columns
$(\lambda_1,\lambda_2,\lambda_3,\lambda_4,\rho_1,\rho_2)$ the four
evaluations give
\begin{equation}
 \begin{pmatrix}
 -16&24&2&12&16&8\\
 0&16&-2&8&16&2\\
 -16&-8&0&-4&-16&-2\\
 -8&-4&1&-6&-8&-2
 \end{pmatrix}.
 \label{affinematrix}
\end{equation}
The first four columns have determinant $-6144$, so the rank is four.
For reference, the reduced row-echelon form is
\begin{equation}
 \begin{pmatrix}
 1&0&0&0&\tfrac12&0\\
 0&1&0&0&1&0\\
 0&0&1&0&0&1\\
 0&0&0&1&0&\tfrac12
 \end{pmatrix}.
 \label{affinematrix-rref}
\end{equation}
Therefore
\begin{equation}
 \lambda_2=2\lambda_1,\qquad \rho_1=-2\lambda_1,
 \qquad \lambda_3=2\lambda_4,\qquad \rho_2=-2\lambda_4.
 \label{affinekernelsolution}
\end{equation}
With $a=\lambda_1$ and $b=\lambda_4$ this gives
$v=a(V_1+2V_2-2U_1)+b(2V_3+V_4-2U_2)$, in agreement with
\eqref{affinetangent-main}.

For reproducibility, use null momenta
$p=(1,-1,0,0)$, $q=(1,0,-1,0)$, $r=(1,1,0,0)$ and
$s=(1,0,0,-1)$, and unit spatial polarizations $e_x,e_y,e_z$. The mode
triples for the rows are
\begin{equation}
 \begin{gathered}
 ((p,e_y),(p,e_y),(q,e_x)),\quad
 ((p,e_y),(p,e_z),(q,e_z)),\\
 ((p,e_y),(r,e_z),(q,e_z)),\quad
 ((p,e_y),(q,e_z),(s,e_x)).
 \end{gathered}
\end{equation}
For a mode $(k,e)$ assign $A_{a,I}=e_a k_I$ and
$B_{ab,I}=(k_ae_b-k_be_a)k_I$, lowering indices with $\eta$, and extract
the coefficient of the product of the three mode parameters. These local
free-solution jets do not require momentum conservation and are not
scattering data.

\section{Scattering conventions and an independent local evaluation}\label{scattering-conventions}

For arbitrary color-valued vector legs $a_i^\mu$ with momenta $k_i$, let
\begin{align}
 f_i^{\mu\nu}&=k_i^\mu a_i^\nu-k_i^\nu a_i^\mu,\\
 K_{ij}^{\mu\nu}&=[a_i^\mu,a_j^\nu]+[a_j^\mu,a_i^\nu],\\
 S_{ij}&=2f_i\cdot f_j,
 \qquad T_{ijk}=f_i\cdot K_{jk}+f_j\cdot K_{ik}+f_k\cdot K_{ij}.
\end{align}
The vertices are the coefficient of $z_1\cdots z_n$ in the corresponding
density and there is no additional $n!$ factor. Let $\mathfrak P(I)$ be the
set of unordered complete pairings of an even label set $I$. Then
\begin{align}
 Y_3(1,2,3)&=-\tfrac12 T_{123},\\
 V_4(I)&=2\sum_{P\in\mathfrak P(I)}\prod_{(i,j)\in P}S_{ij},\\
 V_5(I)&=4\sum_{\substack{J\subset I\\|J|=2}}S_JT_{I\setminus J},\\
 W_6(I)&=-96\sum_{P\in\mathfrak P(I)}\prod_{(i,j)\in P}S_{ij},\\
 W_7(I)&=-192\sum_{\substack{J\subset I\\|J|=3}}T_J
                   \sum_{P\in\mathfrak P(I\setminus J)}
                                     \prod_{(i,j)\in P}S_{ij}.
 \label{vertex-rules}
\end{align}
The size of $I$ is the vertex valence in each line. In $S_J,T_J$ the
members of $J$ supply the indices. Internal legs may carry a general
color-valued polarization and these formulas remain multilinear in it.

For a leaf vertex $V$ on external subset $I$, set $p_I=\sum_{i\in I}k_i$.
The current inserted in the adjacent vertex is the leg with momentum
$p_I$ and components
\begin{equation}
 J^{a\mu}_V(I)=-\frac{\eta^{\mu\mu}}{p_I^2}
       V\bigl(\{i:i\in I\},(-p_I,e_a\otimes e_\mu)\bigr),
 \label{leafcurrent}
\end{equation}
with no sum on $\mu$. Summing the remaining central vertex over the
partitions counted in section~\ref{ward} generates each labeled diagram once.
At the physical data used above, $p_I\cdot J^a_V(I)=0$ for every leaf
$Y_3$ and $V_4$ current, including after a Ward replacement. Hence
longitudinal propagator terms give zero.

To check the sign convention, use plane waves $e^{ikx}$ and the Feynman
gauge quadratic action, whose propagator is $-i\eta^{\mu\nu}/p^2$.
A tree with $I$ internal edges has $I+1$ vertices and, at $gh^2$ with
seven external legs, $5+2I$ derivatives. Its phase is therefore
\begin{equation}
 i^{I+1}(-i)^I i^{5+2I}=-(-1)^I.
\end{equation}
Removing the common minus sign leaves exactly one negative real
propagator per edge, as used above. The independent component expansion
works in the commuting polynomial ring with $z_i^2=0$; quotienting by
these squares preserves the desired multilinear coefficient.

\subsection{A short evaluation of the local Ward polynomial}

The nonzero value can also be obtained directly without
enumerating diagrams. We write $f_i=k_i\wedge\varepsilon_i$ without color,
and for $i=1,\ldots,5$ define
\begin{align}
 D_i&=f_6\cdot(k_7\wedge\varepsilon_i)
                   -f_i\cdot(k_7\wedge\varepsilon_6),\\
 M_i&=\sum_{\{(a,b),(c,d)\}\,\text{pairing of }\{1,\ldots,5\}\setminus\{i\}}
                    (f_a\cdot f_b)(f_c\cdot f_d).
\end{align}
There are three pairings in $M_i$. The color assignment implies
\begin{equation}
 \Ward_{7,\trunc}^{[1,2]}=768\sum_{i=1}^{5}D_iM_i,
 \qquad
 f_i\cdot f_j=2\{(k_i\cdot k_j)(\varepsilon_i\cdot\varepsilon_j)
              -(k_i\cdot\varepsilon_j)(\varepsilon_i\cdot k_j)\}.
 \label{hand-certificate}
\end{equation}
\begin{samepage}
At the displayed point,
\begin{center}
\begin{tabular}{crrr}
\toprule
$i$ & $D_i$ & $M_i$ & $768D_iM_i$ \\
\midrule
1 & $0$ & $-46\,080$ & $0$ \\
2 & $0$ & $17\,280$ & $0$ \\
3 & $0$ & $172\,800$ & $0$ \\
4 & $-32$ & $-153\,600$ & $3\,774\,873\,600$ \\
5 & $-96$ & $13\,056$ & $-962\,592\,768$ \\
\bottomrule
\end{tabular}
\end{center}
Their sum is $2\,812\,280\,832$, exactly the full exchange result.
\end{samepage}
The nonzero polynomial value at this nonsingular physical point also
persists under sufficiently small allowed changes of the kinematics and
polarizations and its existence does not require a factorization or soft limit.

\bibliographystyle{unsrt}
\bibliography{references_deser_submission}
\end{document}